# An Economic-based Resource Management and Scheduling for Grid Computing Applications

G. Murugesan[1], Dr.C.Chellappan[2]

[1] Research Scholar, Department of Computer Science and Engineering,
Anna University, Chennai-600 025, Tamil Nadu, India

[2] Professor, Department of Computer Science and Engineering,
Anna University, Chennai-600 025, Tamil Nadu, India

**Abstract**
Resource management and scheduling plays a crucial role in achieving high utilization of resources in grid computing environments. Due to heterogeneity of resources, scheduling an application is significantly complicated and challenging task in grid system. Most of the researches in this area are mainly focused on to improve the performance of the grid system. There were some allocation model has been proposed based on divisible load theory with different type of workloads and a single originating processor. In this paper we introduce a new resource allocation model with multiple load originating processors as an economic model. Solutions for an optimal allocation of fraction of loads to nodes obtained to minimize the cost of the grid users via linear programming approach. It is found that the resource allocation model can efficiently and effectively allocate workloads to proper resources. Experimental results showed that the proposed model obtained the better solution in terms of cost and time.
.

## 1. Introduction

One of the most complicated task in Grid computing is the allocation of resources for a process; ie., mapping of jobs to various resources. This may be a NP-Complete (Non-deterministic Polynomial time) problem. For example, mapping of 50 jobs into 10 resources produces $10^{50}$ possible mappings. This is because every job can be mapped to any of the resources. In our case the allocation is in terms of co-allocation which means that the job is executed on a number of resources instead of single resource. Here resource means processors which are involved in the scheduling process. We used resources and processors simultaneously. The other complexity of resource allocation is the lack of accurate information about the status of the resources. Load balancing and scheduling play a crucial role in achieving utilization of resources in grid environments [20].

Much of the work was done on finding an optimal allocation of resources in Grid computing environments. The scheduling schemes are divided into two main categories; conventional and economical. The conventional strategies consider the overall performance of the system as a metric for determining the system quality. It does not take the cost as factor for scheduling jobs on resources and treat all resources as the same at all. Some examples are SmartNet, AppleS Project, Condor-G, NetSolve etc. In economic strategy, cost is considered as essential factor for scheduling jobs. The user is charged based on the utility of the resources in the Grid system. Some of the works consider the economic strategies which deals with the price of resources when it needs to allocate jobs to resources and that price usually reflects the value of the resource to the user.

Task scheduling is an integrated part of parallel and distributed computing. The Grid scheduling is responsible for resource discovery, resources selection, job assignment and aggregation of group of resources over a decentralized heterogeneous system; the resources belong to multiple administrative domains. The resources are requested by a Grid application, which use to computing, data and network resources etc. However, Scheduling an applications of a Grid system is absolutely more complex than scheduling an applications of a single computer. Because to get the resources information of single computer and scheduling is easy, such as CPU frequency, number of CPU's in a machine, memory size, memory configuration and network bandwidth and other resources connected in the system. But Grid environment is dynamic resources sharing and distributing. Then an application is hard to get resources information, such as CPU load, available memory, available network capacity etc. And Grid environment also hard to classify jobs characteristic, that run in Grid. There are basically two approaches to solve this problems, the first is based on jobs characteristic and second is based on a distributed resources discovery and allocation system. It should optimize the allocation of a job allowing the execution on the optimization of resources. The scheduling in Grid environment has to satisfy a number of constraints on different problems.

The existing scheduler used in TeraGrid and other notable computing grids are dedicated for research purpose and based on deadlines, resource availability, and the description of resources required by a job. Because TeraGrid and most other grid computing implementations are not employed in profit seeking firms, the dollar cost of running jobs is not taken into account in arriving at the optimal schedule. Instead, the scheduling is evolved by matching resources and applications with the objective of improving the hardware performance. As grid computing migrates from scientific to business uses, the allocation of workloads to resources to meet business objectives, such as overall time, cost, or revenue becomes important aspect. This paper introduces a novel frame work for economic scheduling in Grid computing using the mathematical model.

This paper is organized as follows: section 2 presents the related works, the resource allocation model is discussed in the section 3, section 4 describes the design of resource allocation, the experimental results for the proposed model and the conclusion are discussed in the section 5 and section 6 respectively.

## 2. Related works

To date several grid scheduling algorithms have been proposed to optimize the overall grid system performance. The study of managing resources in the Grid environment started from 1960s. The economic problem [14] results from having different ways for using the available resource, so how to decide what is the best way to use them. Utilization of Grid must be cheaper for the Grid user than purchasing their own resources [10] and must satisfy their requirements. On the other hand, resource providers must know if it is worth to provide their resources for usage by the Grid users. Also the pricing of resources should not be per time unit or slot (eg. cost per minutes) [9]. Because it leads to big difference in speeds of processors, so the price per unit time of a processor might cheaper, but the user must pay large amount of money due to slow processing resources. Moreover the users have to know how many time units they need to finish their jobs. Thus the cost of Grid user must be determined based on the tasks the resource is processing.

Deadline scheduling algorithm [18] is one of the algorithms which follow the economic strategy. In aim of this algorithm, to decrease the number of jobs that doesn't meet their deadlines. The resources are priced according to their performance. This algorithm also has a facility of fallback mechanism; which can inform the grid user to resubmit the jobs again, the jobs which are not met the deadline of the available resources. Nimrod/G [3, 4] includes four scheduling algorithms which are cost, time, conservative time and cost-time. Cost scheduling algorithm tries to decrease the amount of money paid for executing the jobs with respect to the deadline. Time scheduling algorithm attempt to minimize the time required to complete the jobs with respect to their budget allotment. The conservative time scheduling algorithm aims to execute the jobs within the stipulated budget and the deadlines. The cost-time scheduling algorithm works as cost scheduling algorithm except that when there are two or more resources with the same price, it employs time scheduling algorithm. It is not dealing with co-allocation.

The effective workload allocation model with single source has been proposed [1] for data grid system. Market-based resource allocation for grid computing [8] supports time and space shared allocations. Furthermore it supports co-allocation. It is supposed that resources have background load which changes with time and that has the highest priority for execution, so they are not fully dedicated to the grid. Gridway [11] is an agent based scheduling system. It aims that to minimize the execution time, total cost and the performance cost ratio of the submitted job. Market economy based resource allocations in Grids [16] are an auction based user scheduling policies for selecting resources were proposed. GridIs [20] a P2P decentralized framework for economic scheduling using tender model. The author tries to perform the process without considering the deadline and the algorithm is implemented with the help of a variable called conservative degree, its value between 0 and 1.

The time and cost trade-off has been proposed [7] with two meta-scheduling heuristics algorithms, that minimize and manage the execution cost and time of user applications. Also they have presented a cost metric to manage the trade-off between the execution cost and time. Compute power market [6] is architecture responsible for managing grid resources, and mapping jobs to suitable resources according to the utility functions used. Parallel virtual machine [2, 17] enables the computational resources to be used as if they are a single high performance machine. It supports both execution on a single and multiple resources by splitting the task into subtasks. Grid scheduling by using mathematical model was proposed [12, 13] with equal portion of load to all the processors; ie., the entire workload received from a source is equally divided and a portion of load is assigned to a processor with the help of random numbers to divide the entire workload from a source. The work closest to ours is [3] where the authors proposed the algorithm that claims to meet budget and deadline constraints of jobs. However, the algorithm proposed is ad-hoc and does not have an explicit objective.

## 3. Resource Allocation Model

A generic grid computing system infrastructure considered here comprises a network of supercomputers and/or a cluster of computers connected by local area networks, as shown in Figure. 1. We consider the problem of scheduling large-volume loads (divisible loads) within a cluster system, which is part of a grid infrastructure.

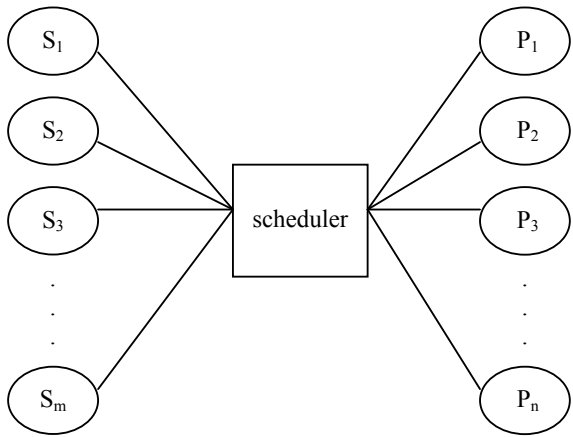

Fig 1: Resource Allocation Model

We envisage this cluster system as a cluster node comprising a set of computing nodes. Communication is assumed to be predominant between such cluster nodes and is assumed to be negligible within a cluster node. The underlying computing system within a cluster can be modeled as a fully connected bipartite graph comprising sources, which have computationally intensive loads to be processed (very many operations are performed on them) and computing elements called sinks, for processing loads (data). To organize and coordinate distributed resources participating in the grid computing environment, we utilize the resource model which contain more than one sources and resources, based on which the uniform and scalable and allocating computational resources are identified.

We assume a star like structure with a scheduler as the central component; acts as a controller to the model. The right side of the scheduler is a set of processors (processing elements) to execute the jobs assigned by the scheduler. The other side of the scheduler are a set of sources; grid users. The scheduler collects the load from all the sources and distribute to a set of processors p1, p2, . . . ,pn. Initially all the loads are held by the scheduler. All the communications are starts from the scheduler and there is no direct communication between the processors. For simplicity here we are assuming that the scheduler is only communicating with processors to distribute the loads and get back the result.  In a dynamic Grid environment, nodes may join or leave frequently, and nodes status may change dynamically. In our model, we are considering the environment as static; i.e. once the node joined in the scheduling process it has to active in the entire processing schedule. But in divisible load theory all the resources has to involve or process all source's workload portion. In this model we are not compelling all the resources to process the entire source's workload portion, instead only the selected set of resources can participate to perform the portion of load from a source. The contribution of this paper is that we propose unequal division of load from a resource with respect to the processor capacity and the availability and the divisible load is assign to a set of processors participated in the scheduling process. Now, we shall formally define the problem that we address.

## 4. Workload Distribution Model

We assume that a star like structure with a set of p-processors (processing elements). The central component is the scheduler/broker. The scheduler collects the loads from all the grid users called as sources and distribute to a set of processors $p_1, p_2, \ldots, p_n$. The sources are the grid users; who need to use the grid system to perform some task. Initially all the loads are held by the scheduler. All the communications are starts from the scheduler and there is no direct communication between the processors. For simplicity here we are assuming that the scheduler is only communicating element with resources/processors to distribute the loads and get back the result.

In our system we are considered that there is no communication delay to submit a portion of load to a processor. Also the result returning time is negligible. All the processor starts processing only after receiving entire workload which is assigned by the scheduler to them with their capacity. Also we assume that processors have independent communication hardware which allows for simultaneous communications and computations on the previously received loads. Additional constraints may be imposed on the processors to the existence of other more urgent computations or maintenance periods the availability of processor $p_i$ may be restricted to some interval $[r_i, d_i]$. Where $r_i$ be the release time of the processor $p_i$ and $d_i$ be the deadline of the processor $p_i$. By such a restriction we mean that computations may take place only in the interval $[r_i, d_i]$. A message with the load may arrive or start arriving before $r_i$. We assume that computations start immediately after the later of the two events: $r_i$ or the load arrival. The computation time must be fit between the later of the above two events.

Here we are assuming that the loads are received from different sources with different capacity and each load become divisible without any restriction. Each loads received from the different sources are divided into set of different tasks (portion of workload). Maximum of one task may be assigned or allotted to a processor. The load portion is depends upon the capacity of the processor. Suppose, if there are m number of sources $\{S_1, S_2, \ldots, S_m\}$ and n number of processors $\{P_1, P_2, \ldots, P_n\}$ and the workloads from each sources become L = $\{L_1, L_2, \ldots, L_m\}$. Where $L_1$ be the total workload received from the source $S_1$ and so on. Each workload $L_i$ may be divided into $T_1, T_2 \ldots$. The load L can be reordered by the scheduler to achieve good performance of the computation. The scheduler splits the workload into tasks and sends them to processors to perform the specified process. Only a set of processors may be used to perform the workload of a source.

We will denote $\alpha_{ij}$ be the size of the task assigned to the processor $p_j$. It is expressed in load units (eg. in bytes). If $\alpha_{ij}=0$ implies that $p_j$ is not in the set of processor selected to perform the process of $i^{th}$ source workload. The total workload from a source is the sum of sizes of the workload parts. ie. $\sum \alpha_{ij} = L_i$. Not only $p_j$ selected by the scheduler, but also the sequence of activating the processors in $p_j$ and the division of load $L_i$ into chunks $\alpha_{ij}$. Here we are considered the processors are unrelated processors, its communication links and start-up time are specific for the task. Similarly the processor computing rates depends on the processor and the task. Let $z_j$ be the communication rate of the link to processor $p_j$ perceived by task $T_j$. Transferring the portion of workload unit $\alpha_{ij}$ to $p_j$ takes $z_j\alpha_{ij}$ time units. Let $t_j$ be the processing rate of processor $p_j$ perceived by task $T_j$. To process $\alpha_{ij}$ portion of load by processor $p_j$ takes $t_j\alpha_{ij}$ time units. Let $c_j$ the processing cost to process the potion of workload $T_j$. The total processing cost to process a portion of workload $\alpha_{ij}$ by the processor $p_j$ becomes $c_j\alpha_{ij}$ cost units.

In this work we analyze the complexity of scheduling the divisible loads $L_1, L_2, \ldots, L_m$ of sizes $T_1, T_2 \ldots$ on n parallel processors $P_1, P_2, \ldots, P_n$ which are interconnected together. We assume that the processors have sufficient memory buffers to store the received portion of workloads and computations. All the processor will start processing immediately after receiving their entire portion of workloads. One processor can process more than one source's portion of workload.

By constructing a schedule the scheduler decides on the sequence of the tasks, the set of processors assigned to each portion of workloads, the sequence of processor activation and the size of the load parts. Our objective is to minimize the usage of the grid user cost. In this paper we assumed that there no separate start-up time for individual processors and there is no fixed cost to utilize the processors. All the processors are dedicated processors. But practically it is not possible, to simplify our model as well as reduce the number of variables and constraints. The following notations are used to formulate the mathematical model

- $c_j$ - Amount to spend to utilize $j^{th}$ processor
- $\alpha_{ij}$ - Portion of workload from $i^{th}$ source to $j^{th}$ processor
- $x_{ij}$ - Binary variable
- $d_i$ - Deadline to complete the $i^{th}$ source job
- $b_i$ - Budget allotted for the $i^{th}$ source job
- $s_j$ - Schedule period of $j^{th}$ processor
- $t_j$ - Time required to perform operation on one unit of job by $j^{th}$ processor
- $\omega_i$ - Total workload of $i^{th}$ source
- $z_j$ - Time taken to transfer a unit of workload to $j^{th}$ processor
- $s_j$ - Scheduled time $i^{th}$ source

Minimize

$$\sum_i \sum_j c_j \alpha_{ij} x_{ij} \qquad \ldots \ldots (1)$$

Subject to

$$\sum_i \sum_j z_j \alpha_{ij} x_{ij} + \sum_i \sum_j t_j \alpha_{ij} x_{ij} \leq d_i \qquad \ldots \ldots (2)$$

$$\sum_i \sum_j z_j \alpha_{ij} x_{ij} + \sum_i \sum_j t_j \alpha_{ij} x_{ij} \leq s_i \quad ; \forall j \quad \ldots \ldots (3)$$

$$\sum_i \sum_j c_j \alpha_{ij} x_{ij} \leq b_i \qquad \ldots \ldots (4)$$

$$\sum_j \alpha_{ij} = \omega_i \quad ; \forall i \quad \ldots \ldots (5)$$

$$\sum_i x_{ij} = 1 \quad ; \forall j \quad \ldots \ldots (6)$$

$$\alpha_{ij} \geq 0 \quad ; \forall i, j \quad \ldots \ldots (7)$$

$$x_{ij} = \{0,1\} \quad ; \forall i, j \quad \ldots \ldots (8)$$

$$s_j \geq 0 \quad ; \forall j \quad \ldots\ldots\ldots (9)$$
$$z_j \geq 0 \quad ; \forall j \quad \ldots\ldots\ldots (10)$$

The objective is to minimize the total cost of the grid user those who are assigning job to the grid system. The equation (1) represent the cost of all jobs that are being assigned to a resource is the objective function. The equation (2) to (8) specifies the constraints used the mathematical model. The constraints (2) represent the deadline associated with each sources. Constraints (3) match the workload within the availability of resources. Constraints (4) are the budget for each source's workload. Constraint (5) represents the total workload of each sources involved in the scheduling process. Constraints (6) makes sure that a portion of workload is assigned to only one resource; ie., there is no overlap between processing of workloads. Constraint (7) makes sure that a portion of workload divided from the total workload becomes a whole number. Constraint (8) is to set the binary value either 0 or 1. The constraints (9 and 10) are non-negativity constraints.

## 5. Experimental results

Let us assume that the Grid system consists of five processors (resources) namely $P_1$, $P_2$, $P_3$, $P_4$, and $P_5$ with four sources $S_1$, $S_2$, and $S_3$ are trying to utilize the grid system to execute their workloads.

Table I : Details of Processors Capacity

| Processor | Processing time/ unit workload (min) | Processing cost (Rs.) | Available time (min.) |
|---|---|---|---|
| $P_1$ | 3 | 4 | 60 |
| $P_2$ | 4 | 3 | 60 |
| $P_3$ | 5 | 2 | 80 |
| $P_4$ | 4 | 3 | 110 |
| $P_5$ | 3 | 5 | 110 |

Table I shows that the processor's involved in the Grid system, the processing capacity of each processor per unit workloads, the processing cost of each processor to execute a unit workload and available time of each processors. Table II shows that the details of the different sources which are trying to utilize the grid system, total workload of each sources and the budget allotted to the each sources to complete their workloads and the expected time to complete the process of each workloads. Using the details given in the table we have formed the mathematical model and solved the equation using LINGO package. After execution of the mathematical model, the maximum cost to spend for processing all the three sources workloads are Rs.1457.

Table II : Details of Workloads of sources, Budget and Finish Time

| Sources | Work Loads (MB) | Budget (Rs) | Dead Line (min.) |
|---|---|---|---|
| $S_1$ | 30 | 120 | 100 |
| $S_2$ | 35 | 135 | 130 |
| $S_3$ | 45 | 180 | 175 |

Table III : Allocation of Processors

| Sources | Processor Allotted | Allotted Workloads | Time taken to complete |
|---|---|---|---|
| $S_1$ | $P_1$ | 18 | 100 |
| | $P_2$ | 10 | |
| | $P_5$ | 2 | |
| $S_2$ | $P_3$ | 7 | 129 |
| | $P_4$ | 10 | |
| | $P_5$ | 18 | |
| $S_3$ | $P_2$ | 5 | 174 |
| | $P_3$ | 9 | |
| | $P_4$ | 16 | |
| | $P_5$ | 15 | |

Table III shows the details of workload allotment to the processors involved in the process. From the table it is clear that the total workload of $S_1$ is divided into three parts, the total workload of $S_2$ is divided into three parts and the total workload of $S_3$ is divided into four parts and allotted into processors. Also it shows that the completion time of each source's workloads

## 6. Conclusion

In this study, we have developed an effective iterative model for optimal workload allocation. The proposed model is proposed for load allocation to processors and links for scheduling divisible workload applications. The experimental results showed that the proposed model is capable of producing almost optimal solution for multiple sources scheduling with static and dedicated resources. Hence the proposed model can balance the processing loads efficiently. We are planning to adapt the proposed model in dynamic environments. With such improvements the proposed model can be integrated in the existing grid scheduler in order to improve their performance.

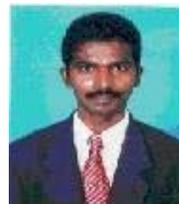

G.Murugesan received the B.E. in Computer Science and Engineering from Manonmanium Sundaranar University, Tirunelveli, Tamil Nadu, India in 1996, the M.E. in Systems Engineering and Operations Research from Anna University, Chennai, Tamil Nadu, India in 2006. Presently pursuing Ph.D. in the area of Grid Computing in Anna University, Chennai.

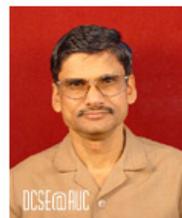

C.Chellappan received his Ph.D in the area of Database Systems from Anna University, Chennai, Tamil Nadu, India where he is currently the Professor of Computer Science and Engineering Department. His research interest includes Mobile Computing, Sensor Networks and Parallel system scheduling. He is also a Principal Investigator of project(Rs One Crore) on "Basic directed Collaborative Research in Smart and  Secure Environment " sponsored by National Technical Research Organization, New Delhi, Govt of India, 2007-2010.